# Critical study and discrimination of different formulations of electromagnetic force density and consequent stress tensors inside matter

Amir M. Jazayeri[*] and Khashayar Mehrany[ł]

*Department of Electrical Engineering, Sharif University of Technology, Tehran 145888-9694, Iran*

By examination of the exerted electromagnetic (EM) force on *boundary* of an *object* in a few examples, we look into the compatibility of the stress tensors corresponding to different formulae of the EM force density with special relativity. Ampere-Lorentz's formula of the EM force density is physically justifiable in that the electric field and the magnetic flux density act on the densities of the *total* charges and the *total* currents, unlike Minkowski's formula which completely excludes the densities of the *bounded* charges and the *bounded* currents inside *homogeneous* media. Abraham's formula is fanciful and devoid of physical meaning. Einstein-Laub's formula seems to include the densities of the *total* charges and the *total* currents at first sight, but grouping the *bounded* charges and the *bounded* currents into point-like dipoles erroneously results in the hidden momentum being omitted, hence the error in [Phys. Rev. Lett. 108, 193901 (2012)]. Naturally, the Ampere-Lorentz stress tensor accords with special relativity. The Minkowski stress tensor is also consistent with special relativity. It is worth noting that the mathematical expression of the Minkowski stress tensor can be quite different from the well-known form of this stress tensor in the literature. We show that the Einstein-Laub stress tensor is incompatible with special relativity, and therefore we rebut the Einstein-Laub force density. Since the Abraham momentum density of the EM fields is inherently corresponding to the Einstein-Laub force density [Phys. Rev. Lett 111, 043602 (2013)], our rebuttal may also shed light on the controversy over the momentum of light.

I.   INTRODUCTION

The exerted electromagnetic (EM) force on a point charge is given by the Lorentz force law which, together with Maxwell's equations, constitutes one of the five independent principles of EM theory. No matter how Maxwell's equations are mathematically formulated [1], they always express the same notion and yield the same results. In contrast, different mathematical expressions that have been hitherto reported for the EM force density within a material medium could yield different results even when they are applied to one particular problem [2-3].

A survey of the literature shows that the most famous expressions of the EM force density are Ampere-Lorentz's, Einstein-Laub's, Minkowski's, and Abraham's formulae [2-10]. The most physically appealing formula is Ampere-Lorentz's. Still, Einstein has been quoted to disapprove his own formula and endorse Minkowski's [8, 11]. More recently, particular attention has been directed toward Ampere-Lorentz's and Einstein-Laub's formulae [4-5, 8-10]. It was mistakenly asserted that the former is inconsistent with special relativity [7]. This wrong assertion was soon rectified in scores of publications [12-16].

The Ampere-Lorentz expression of the EM force density is the only one physically justifiable because the electric field ($\vec{E}$) and the magnetic flux density ($\vec{B}$) act on the densities of the *total* charges and the *total* currents. Einstein-Laub's formula seems to have such property, but grouping pairs of charges into point-like dipoles results in the hidden momentum being omitted. The lack of the hidden momentum is exactly the same error recently made in [7]. Minkowski's and Abraham's formulae are physically unjustifiable in that they completely exclude the densities of the *bounded* charges and the *bounded* currents inside *homogeneous* media. Moreover, Abraham's formula is based on an esthetic consideration which is devoid of physical meaning.

Once either one of the aforementioned formulae is adopted for the EM force density within a material medium, the overall EM force exerted upon the entire volume of the material can be easily obtained by volume integration of the EM force density. This integral is usually written as a summation of two integrals; first, surface integral of the stress tensor, and second, time rate of decrease of the volume integral representing the momentum of the EM fields inside the volume [2-3]. Obviously, different formulations result in different stress tensors and momentum densities. Ampere-Lorentz's force density leads to the Ampere-Lorentz stress tensor and the Livens momentum density. Einstein-Laub's force density leads to the Einstein-Laub stress tensor and the Abraham momentum density. Minkowski's force density leads to the Minkowski stress tensor and the Minkowski momentum density. Abraham's force density leads to the Abraham stress tensor and the Abraham momentum density.

Naturally, the debate on the correct form of the EM force density within a material body goes in tandem with the debate on the correct form of the EM stress tensor and the EM momentum density within a material body. Nevertheless, finding the correct form of the EM stress tensor in material medium has not been a widely studied subject in the literature. Rather, the Minkowski stress tensor, being referred to as the Maxwell stress tensor in Jackson's book [17], other texts [18], and research papers [19-23], has been the most common stress tensor in the literature. The correct form of the momentum density in a material medium, on the contrary, has been the subject of controversy in the past [24-28].

In this manuscript, we *theoretically* assess different formulations of the EM stress tensor and thereby evaluate different formulations of the EM force density. To this end, the EM force exerted upon the *boundary* of material *object*s is first calculated by using different stress tensors. In view of the fact that the method of virtual work cannot be accounted a solid criterion for the correct EM force, the compatibility of the obtained results with special relativity is used as the *theoretical* criterion. Ironically, it is shown that the Einstein-Laub formulation is not necessarily consistent with Einstein's special relativity. This observation is in stark contrast to the assertion that the Einstein-Laub formulation is in perfect harmony with special relativity [7].

The organization of this paper is as follows. First, different formulations of the EM stress tensor are reviewed in Section II. Since the mathematical expressions of the stress tensors of Minkowski and Abraham inside a material depend on the constitutive relations that govern the displacement field ($\vec{D}$) and the magnetic flux density ($\vec{B}$), we specifically derive these tensors for three categories of *homogeneous* materials: linear isotropic dispersive/non-dispersive materials, linear reciprocal non-dispersive materials with permittivity and/or permeability tensors, and nonlinear materials with permanent electric and/or magnetic dipoles. Peculiarly, the forms of the stress tensors in the latter category of materials are quite different from the well-known forms of the stress tensors used in the literature. This is also in contrast to the assertion of some authors that the applicability of the Minkowski force density is limited to linear materials [8]. In Section III, from a viewpoint based on the law of conservation of momentum in EM interactions, the generic problem of the EM force experienced by an *object* is addressed. Thereby, the total EM force exerted on the *object* is found together with the distribution of the exerted EM force on the *object boundary*. Despite the opinion of some authors [22], we argue that finding the exact amount of the total EM force exerted on the *object* in a non-free space host medium requires more than a macroscopic study of the interface. Besides, all the aforementioned formulations of the EM force density render the same total EM force (or time-averaged EM force) exerted on the *object* in free space only when the EM fields are static (or time-harmonic). It should be noted that even though the contribution of the Helmholtz term in the Minkowski (or Abraham) force density [2-3, 29] is no longer zero because there is a discontinuity on the *object boundary*, finding and integrating the contribution of the Helmholtz term in the exerted Minkowski (or Abraham) force is sidestepped by following the approach of this section. Section IV is devoted to a *theoretical* assessment of the different formulations of the EM stress tensor. Although the difference among the results obtained by using different formulations for the EM force density is an already known fact [8, 30], no conclusive experiment has been carried out to decide on the correct formulation. Regretfully, the result of the method of virtual work cannot be accounted a solid criterion for the correct EM force because the resulting EM force depends on the constitutive relations within the material *object* in its *virtually expanded/contracted* state. Therefore, the compatibility of the obtained results with special relativity is exploited as a solid *theoretical* criterion. Finally, concluding remarks are provided in Section V.

## II. DIFFERENT FORMULATIONS OF EM FORCE DISTRIBUTION

### A. The formulations of Ampere-Lorentz and Einstein-Laub

Ampere and Lorentz naturally, and obviously from the physical point of view, considered that $\vec{E}$ and $\vec{B}$ act on the densities of the *total* charges and the *total* currents, i.e. not only the density of the *free* charges $(\rho)$ and the density of the *free* currents $(\vec{J})$, but also the densities of the *bounded* charges $(-\nabla \cdot \vec{P})$ and the *bounded* currents $(\frac{\partial \vec{P}}{\partial t} + \nabla \times \vec{M})$. Therefore, the Ampere-Lorentz force density reads as:

$$\vec{f}_{Am} = (\rho - \nabla \cdot \vec{P})\vec{E} + (\vec{J} + \frac{\partial \vec{P}}{\partial t} + \nabla \times \vec{M}) \times \vec{B}. \tag{1}$$

The Einstein-Laub force density, on the other hand, is as follows:

$$\vec{f}_E = (\rho + \vec{P}\cdot\nabla)\vec{E} + \mu_0(\vec{J} + \frac{\partial\vec{P}}{\partial t})\times\vec{H}$$

$$+ \mu_0(\vec{M}\cdot\nabla)\vec{H} - \varepsilon_0\mu_0\frac{\partial\vec{M}}{\partial t}\times\vec{E}. \tag{2}$$

According to [4], the difference between the Ampere-Lorentz and the Einstein-Laub force densities in a nonmagnetic medium ($\vec{M} = 0$) is that the macroscopic electric polarization ($\vec{P}$) is modeled as an aggregate of microscopic charges and microscopic dipoles for the former and latter cases, respectively. It was just the grouping of pairs of separated charges into point-like dipoles that led Einstein and Laub to an error, because separated charges considered at the same time in their rest frame are subject to forces at different times in a moving frame. The grouping of the *bounded* charges and the *bounded* currents into point-like dipoles erroneously results in the hidden momentum, which is the necessary quantity arising from non-conservation of simultaneity of separate events, being omitted. This is exactly the error made in [7]. Although the error had not been discovered at the Einstein-Laub times, there are papers published decades ago on the subject of relativistic treatment of extended bodies in general [31-33], and magnetic dipoles in particular [34-35].

It can be seen that both of the formulae lead to $(\vec{p}_0\cdot\nabla)\vec{E} + \mu_0(\partial\vec{p}_0/\partial t)\times\vec{H}$ as the exerted force on a point-like electric dipole with the electric polarization $\vec{P} = \vec{p}_0\delta(\vec{r})$. The exerted force on a point-like magnetic dipole with the magnetic polarization $\vec{M} = \vec{m}_0\delta(\vec{r})$ is $\mu_0(\vec{m}_0\cdot\nabla)\vec{H} + \mu_0\varepsilon_0\vec{m}_0\times(\partial\vec{E}/\partial t)$, when Ampere-Lorentz's formula is applied, and $\mu_0(\vec{m}_0\cdot\nabla)\vec{H} - \mu_0\varepsilon_0(\partial\vec{m}_0/\partial t)\times\vec{E}$, when Einstein-Laub's formula is applied. The difference between the obtained forces equals time rate of change of the hidden momentum, i.e. $\partial(\vec{m}_0\times\vec{E})/(c^2\partial t)$.

Integrations of the Ampere-Lorentz and the Einstein–Laub force densities over an arbitrary volume $\upsilon$, which is *not* necessarily *homogenous*, result in a summation of a volume integral of time rate of decrease of the momentum density $\vec{G}$, and a surface integral of the stress tensor $\ddot{T}$:

$$\int_{\upsilon} \vec{f} d\upsilon = \int_{\upsilon} \frac{-\partial \vec{G}}{\partial t} d\upsilon + \oint_{\partial \upsilon} \hat{n} \cdot \ddot{T} dS, \tag{3}$$

where $\partial \upsilon$ is the boundary of $\upsilon$, $\hat{n}$ is the unit vector normal to the boundary element $dS$ toward the outside of $\upsilon$. For the Ampere-Lorentz force density, we have the Livens momentum density $\vec{G}_L = \varepsilon_0 \vec{E} \times \vec{B}$, and the Ampere-Lorentz stress tensor $\ddot{T}_{Am}$ [2-3]:

$$\ddot{T}_{Am} = \varepsilon_0 \vec{E}\vec{E} + \frac{1}{\mu_0} \vec{B}\vec{B} - \frac{1}{2}\ddot{I}(\varepsilon_0 \vec{E} \bullet \vec{E} + \frac{1}{\mu_0} \vec{B} \bullet \vec{B}). \tag{4}$$

For the Einstein-Laub force density, we have the Abraham momentum density $\vec{G}_{Ab} = \varepsilon_0 \mu_0 \vec{E} \times \vec{H}$, and the Einstein-Laub stress tensor $\ddot{T}_E$ [2-3]:

$$\ddot{T}_E = \vec{D}\vec{E} + \vec{B}\vec{H} - \frac{1}{2}\ddot{I}(\varepsilon_0 \vec{E} \cdot \vec{E} + \mu_0 \vec{H} \cdot \vec{H}). \tag{5}$$

It is worth noting that, unlike the stress tensors of Minkowski and Abraham, the forms of the stress tensors of Ampere-Lorentz and Einstein-Laub in Eqs. (4) and (5) are independent of the constitutive relations of $\vec{D}$ and $\vec{B}$.

### B. The formulations of Minkowski and Abraham

The Minkowski force density seems to be physically unjustifiable in that it completely excludes the densities of the *bounded* charges and the *bounded* currents inside *homogeneous* media. The Abraham force density is fanciful because it is based on the Minkowski force density together with a merely esthetic consideration which is devoid of physical meaning.

In this subsection, the Minkowski stress tensor is derived by integration of the Minkowski force density over an arbitrary volume within *homogeneous* media. The importance of this derivation lies in the fact that the Minkowski formulation is not limited to linear materials. This is in contrast to what is explicitly stated in some papers [8].

The Minkowski force density, inside a *homogeneous* medium, is written as

$$\vec{f}_M = \rho\vec{E} + \vec{J} \times \vec{B},  \quad (6)$$

where $\rho$ and $\vec{J}$ are the densities of the *free* charges and the *free* currents, respectively. In *inhomogeneous* media, a Helmholtz term has to be included in the Minkowski force density, which for linear and isotropic media appears as $-(\vec{E}\cdot\vec{E}/2)\nabla\varepsilon - (\vec{H}\cdot\vec{H}/2)\nabla\mu$ [2-3, 29]. Despite the opinion of some authors [8], the Minkowski formulation is not limited to linear materials.

As already mentioned, every formula of the EM force density can be integrated over an arbitrary volume $\upsilon$ to obtain the associated EM force exerted on the volume $\upsilon$. It is shown in Appendix A that the integration of Eq. (6) over the volume $\upsilon$ within a *homogeneous* medium is analogous to the expression in Eq. (3) and therefore includes a momentum density and a stress tensor. $\vec{G}_M = \vec{D} \times \vec{B}$ is the Minkowski momentum density of the EM fields inside the medium, and $\vec{\vec{T}}_M$ is the Minkowski stress tensor which reads as

$$\vec{\vec{T}}_M = \vec{D}\vec{E} + \vec{B}\vec{H} - \frac{1}{2}\vec{\vec{I}}(\vec{D}\cdot\vec{E} + \vec{B}\cdot\vec{H}), \quad (7)$$

for a linear reciprocal medium and

$$\vec{\vec{T}}_M = \vec{D}\vec{E} + \vec{B}\vec{H} - \frac{1}{2}\vec{\vec{I}}(\varepsilon_0\vec{E}\cdot\vec{E} + 2\vec{P}_0\cdot\vec{E} + \vec{B}\cdot\vec{H}), \quad (8)$$

for a nonlinear medium with a permanent electric polarization $\vec{P}_0$ and a linear reciprocal magnetic response. In the same manner, the Minkowski stress tensor for other types of *homogeneous* media, e.g. a nonlinear medium with a permanent magnetization $\vec{M}_0$, can be derived.

The Minkowski stress tensor is not symmetric. The Abraham force density is obtained in such a manner that its corresponding stress tensor becomes symmetric, and its corresponding momentum density becomes the Abraham momentum density $\vec{G}_{Ab} = \varepsilon_0 \mu_0 \vec{E} \times \vec{H}$. It can be shown that the Abraham force density inside a *homogeneous* medium is as follows:

$$\vec{f}_{Ab} = \vec{f}_M + \frac{\partial(\vec{G}_M - \vec{G}_{Ab})}{\partial t}$$

$$-\frac{1}{2}\sum_{i=x,y,z}\frac{\partial(\vec{E}D_i)}{\partial i} + \frac{1}{2}\sum_{i=x,y,z}\frac{\partial(\vec{D}E_i)}{\partial i}. \qquad (9)$$

The previous remark about the absence of the Helmholtz term for *homogeneous* media remains true for the Abraham force density as well. The Abraham counterparts of the Minkowski stress tensors in Eq. (7), i.e. for a linear reciprocal medium, and Eq. (8), i.e. for a nonlinear medium with a permanent electric polarization $\vec{P}_0$ and a linear reciprocal magnetic response, can be obtained by following the same steps. It can be shown that these counterparts are

$$\vec{\vec{T}}_{Ab} = \frac{1}{2}(\vec{D}\vec{E} + \vec{E}\vec{D} + \vec{B}\vec{H} + \vec{H}\vec{B}) - \frac{1}{2}\vec{\vec{I}}(\vec{D}\cdot\vec{E} + \vec{B}\cdot\vec{H}), \qquad (10)$$

and

$$\vec{\vec{T}}_{Ab} = \frac{1}{2}(\vec{D}\vec{E} + \vec{E}\vec{D} + \vec{B}\vec{H} + \vec{H}\vec{B})$$

$$-\frac{1}{2}\vec{\vec{I}}(\varepsilon_0\vec{E}\cdot\vec{E} + 2\vec{P}_0\cdot\vec{E} + \vec{B}\cdot\vec{H}), \qquad (11)$$

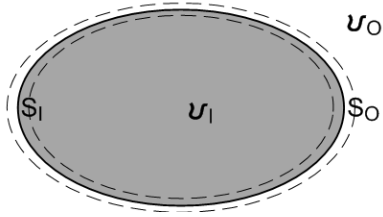

Fig. 1. Schematic of a *homogeneous object* placed within a *homogeneous* host medium

respectively. It is worth noting that, for a linear isotropic material, the stress tensors of Minkowski and Abraham are the same.

### III. EM FORCE EXPERIENCED BY AN OBJECT

In this section, we investigate how the EM force exerted on an *object* is distributed over its *boundary*. Fig. 1 schematically shows a typical situation when the *object* is placed within a host medium. In this figure, $\upsilon_I$, $\upsilon_O$, and $S_B$ denote the region inside the *object*, the region outside the *object*, and the *object boundary*, respectively. It is worth noting that the *object boundary* belongs to both the *object* and the host medium. Obviously, the exact treatment of the interface requires a microscopic study of the interface between inside and outside the *object*. However, the macroscopic constitutive relations, e.g. electric permittivity or magnetic permeability, become meaningless in the microscopic study of the interface. In the here adopted macroscopic point of view, the necessity of going through a rather meticulous microscopic study of the interface is sidestepped by approaching the interface region once from within the *object* and once from without the *object*. Therefore, we consider $S_I$ and $S_O$, which are imaginary surfaces that approach the *object boundary* from $\upsilon_I$ and $\upsilon_O$, respectively.

#### A. The Ampere-Lorentz force and the Einstein-Laub force

The Ampere-Lorentz (or Einstein-Laub) force experienced by an *object* inside a host medium is easy to obtain because the Ampere-Lorentz (or Einstein-Laub) stress tensor in Eq. (4) (or Eq. (5)), which is resulting from integration of the Ampere-Lorentz (or Einstein-Laub) force density, is general and independent of whether the integration volume is *homogeneous*. Therefore, the followings are the exerted Ampere-Lorentz force and the Einstein-Laub force on the *object boundary* $S_B$, respectively:

$$\oint_{S_B} \hat{n} \cdot (\vec{\vec{T}}_{Am_O} - \vec{\vec{T}}_{Am_I}) dS, \tag{12}$$

$$\oint_{S_B} \hat{n} \cdot (\vec{\vec{T}}_{E_O} - \vec{\vec{T}}_{E_I}) dS, \tag{13}$$

where $\vec{\vec{T}}_{Am_I}$ (or $\vec{\vec{T}}_{E_I}$) and $\vec{\vec{T}}_{Am_O}$ (or $\vec{\vec{T}}_{E_O}$) are the Ampere-Lorentz (or Einstein-Laub) stress tensors inside $\upsilon_I$ and $\upsilon_O$, respectively, and $\hat{n}$ is the unit vector normal to the surface element $dS$ toward the outside of $\upsilon_I$. Since the total Ampere-Lorentz (or Einstein-Laub) force exerted on the *object boundary* $S_B$ is written as a surface integral, its integrand is the distribution of the Ampere-Lorentz (or Einstein-Laub) force exerted on the *object boundary* $S_B$.

It is worth noting that the *object boundary* $S_B$ in our treatment is in fact the macroscopic representation of the microscopic volume in between the *object* and the host medium. Since the microscopic volume has a non-zero mass, attributing a non-zero force to the two-dimensional *object boundary* $S_B$ is not paradoxical.

The total Ampere-Lorentz force and the total Einstein-Laub force exerted on $\upsilon_I$ and $S_B$ can be written as

$$\int_{\upsilon_I} \frac{-\partial \vec{G}_L}{\partial t} d\upsilon + \oint_{S_B} \hat{n} \cdot \vec{\vec{T}}_{Am_O} dS, \tag{14}$$

$$\int_{\upsilon_I} \frac{-\partial \vec{G}_{Ab}}{\partial t} d\upsilon + \oint_{S_B} \hat{n} \cdot \vec{\vec{T}}_{E_O} dS, \tag{15}$$

respectively, where $\hat{n}$ is the unit vector normal to the surface element $dS$ toward the outside of $\upsilon_I$. Regretfully, the macroscopic point of view does not clarify how much of Eq. (14) (or Eq. (15)) contributes to acceleration of the *object* center of mass because the *object boundary* $S_B$ in principle belongs to both the *object* and the host medium. Nevertheless, there is no ambiguity about the total Ampere-Lorentz (or Einstein-Laub) force that accelerates the *object* center of mass when the host medium is devoid of matter, i.e. free space. In such a case, the *object boundary* $S_B$ is not shared between two media (the *object* and its host medium) but belongs to the *object* totally. Therefore, Eq. (14) (or Eq. (15)) becomes the overall Ampere-Lorentz (or Einstein-Laub) force exerted on the *object*.

### B. The Minkowski force and the Abraham force

Finding the Minkowski (or Abraham) force experienced by an *object* inside a host medium is a more delicate process because the form of the Minkowski (or Abraham) stress tensor depends on the constitutive relations. Our derivation of the Minkowski (or Abraham) force experienced by the *object* is based on the assumption that the *object* and its host medium are both made of *homogeneous* materials. The distributions of the *free* charges and the *free* currents are arbitrary.

Now, according to the results of Section II-B the Minkowski force exerted on $\upsilon_I$, as a *homogenous* medium, is written as

$$\int_{\upsilon_I} \frac{-\partial \vec{G}_M}{\partial t} d\upsilon + \oint_{S_I} \hat{n} \cdot \ddot{\vec{T}}_M dS, \qquad (16)$$

where $\hat{n}$ is the unit vector normal to the surface element $dS$ toward the outside of $\upsilon_I$. In a similar fashion, the Minkowski force exerted on $\upsilon_O$, as the other *homogeneous* medium, is written as

$$\int_{\upsilon_O} \frac{-\partial \vec{G}_M}{\partial t} d\upsilon + \oint_{S_O} \hat{n} \cdot \vec{\vec{T}}_M dS, \tag{17}$$

where $\hat{n}$ is the unit vector normal to the surface element $dS$ toward the outside of $\upsilon_O$. On the other hand, according to the law of conservation of momentum in EM interactions, the total Minkowski force exerted on the whole system, which comprises $\upsilon_I$, $\upsilon_O$, and $S_B$, should equal time rate of decrease of the total momentum of the EM fields. Comparing the summation of the Minkowski forces exerted on $\upsilon_I$, and $\upsilon_O$, i.e. the summation of Eqs. (16) and (17) when the imaginary surfaces $S_I$ and $S_O$ approach the real interface $S_B$, against time rate of decrease of the total Minkowski momentum of the EM fields reveals that there should be a Minkowski force exerted on the *object boundary* $S_B$:

$$\oint_{S_B} \hat{n} \cdot (\vec{\vec{T}}_{M_O} - \vec{\vec{T}}_{M_I}) dS, \tag{18}$$

where $\vec{\vec{T}}_{M_I}$ and $\vec{\vec{T}}_{M_O}$ are the Minkowski stress tensors inside $\upsilon_I$ and $\upsilon_O$, respectively, and $\hat{n}$ is the unit vector normal to the surface element $dS$ toward the outside of $\upsilon_I$. The integrand in Eq. (18) is the distribution of the Minkowski force exerted on the *object boundary* $S_B$. This treatment of the *object boundary* is clearly distinct from the approach based on the inclusion of the Helmholtz term, which usually appears as $-(\vec{E} \cdot \vec{E}/2)\nabla \varepsilon - (\vec{H} \cdot \vec{H}/2)\nabla \mu$ in the Minkowski force density [2-3, 29]. Interestingly, the line integral of the Helmholtz term over an infinitesimal path normal to $S_B$ is equivalent to $\hat{n} \cdot (\vec{\vec{T}}_{M_O} - \vec{\vec{T}}_{M_I})$, where $\hat{n}$ is the local unit vector normal to $S_B$ toward the outside of $\upsilon_I$.

Now, the total Minkowski force exerted on $\upsilon_I$ and $S_B$ can be written as summation of Eqs. (16) and (18), which is

$$\int_{\upsilon_I} \frac{-\partial \vec{G}_M}{\partial t} d\upsilon + \oint_{S_B} \hat{n} \cdot \vec{\vec{T}}_{M_O} dS, \qquad (19)$$

where $\hat{n}$ is the unit vector normal to the surface element $dS$ toward the outside of $\upsilon_I$. Since the *object boundary* $S_B$ is shared between the *object* and its host medium, finding the overall Minkowski force exerted on the *object*- much like finding the overall Ampere-Lorentz and Einstein-Laub forces- requires more than a macroscopic study of the interface unless the host medium is free space. In such a case, Eq. (19) becomes the overall Minkowski force exerted on the *object*. Therefore, despite the opinion of some authors [22], neither Eq. (19) nor its equivalent forms for static and time-harmonic EM fields [36] can express the overall Minkowski force exerted on the *object* when the host medium is not free space.

In a similar fashion, the exerted Abraham forces on $\upsilon_I$, $\upsilon_O$, and $S_B$ can be obtained. It can be easily shown that the Abraham counterpart of Eq. (18) reads as

$$\oint_{S_B} \hat{n} \cdot (\vec{\vec{T}}_{Ab_O} - \vec{\vec{T}}_{Ab_I}) dS. \qquad (20)$$

If the *object* and the host medium are both linear and isotropic, Eqs (20) and (18) become equivalent to each other and therefore the formulations of Abraham and Minkowski render the same distribution for the exerted EM force on the *object boundary*.

Along the same line, the overall Abraham force exerted on the *object* in free space is obtained by replacing $\vec{G}_M$ and $\vec{\vec{T}}_{M_O}$ with $\vec{G}_{Ab}$ and $\vec{\vec{T}}_{Ab_O}$, respectively in Eq. (19).

Interestingly, the overall Ampere-Lorentz, Einstein-Laub, Minkowski, and Abraham forces exerted on the *object* in free space when the EM fields are static (or their time averages when the EM fields are time-harmonic) become equal to each other.

IV.     ASSESSMENT OF FORMULATIONS OF EM FORCE DISTRIBUTION

Compatibility with special relativity is a solid *theoretical* criterion for correct formulation of the EM stress tensor and its corresponding force density. Some tried to show that the Ampere-Lorentz force density is inconsistent with special relativity [7]. But this assertion was incorrect because the Ampere-Lorentz force density as a generalization of the Lorentz force law has a covariant 4-vector [12, 13, 37], and the Ampere-Lorentz stress tensor has a covariant 4-tensor [37]. From another point of view adopted by some other authors [14, 15], the omission of the hidden momentum [38-40] leads to the wrong result reported in [7].

It is worth noting that one can discard the hidden momentum by employing the asynchronous formulation instead of the synchronous formulation [31-33]. According to the synchronous formulation, the condition of equilibrium in an inertial frame is met when sum of the forces (and torques) exerted on the constituent parts of the extended body under investigation is synchronously zero. In the asynchronous treatment; however, the forces (and torques) exerted on the constituent parts are to be considered asynchronously, i.e. at different times. In this latter approach, there is only one inertial frame in which the forces (and torques) are summed at the same time.

If the EM fields are static (or time-harmonic) in both the rest frame and a moving inertial frame, we don't have to be worried about the hidden momentum insofar as the EM force (or the time-averaged EM force) is considered. In such a case, there is no need to resort to the asynchronous formulation, and the results of the previous sections for static (or time-harmonic)

EM fields are applicable not only in the rest frame but also in the moving frame. In contrast, insofar as torque is considered, even if the forces are static in both the rest and the moving frames, either the hidden entities should be included, or the asynchronous formulation is to be adopted. This is due to the fact that torque is the cross product of the position vector and force, and the positions of the constituent parts vary with time in the moving frame. One famous example is the right-angled lever problem [33].

Here, two examples, named Example A and Example B, are devised specifically to rebut the assertion that the Einstein-Laub formulation is in harmony with special relativity [7]. The examples are static in not only the rest frame but also in a moving frame. It is worth noting that if the EM fields are time-harmonic in the rest frame, they are time-harmonic in any other inertial frame. This is not necessarily the case for static EM fields. That is why we emphasize that our examples are static in both frames.

In each of the examples, we examine the EM force experienced by a linear isotropic *homogeneous object* with a permittivity $\varepsilon$ and a permeability $\mu$ inside free space in the presence of the *free* charges. For the sake of brevity, the counterparts of the examples in the presence of the *free* currents are only mentioned en passant. From the remarks made in the previous section it is evident that; first, all the formulations of the EM force density agree on the total EM force exerted on the *object* in each of the examples, and therefore, the discussion focuses on the distribution of the exerted EM force on the *object boundary*; and second, the results of the formulations of Minkowski and Abraham are identical in each of the examples.

Due to the static nature of our examples, one might be tempted to call upon the method of virtual work rather than special relativity to find the correct force distribution. The method of virtual work, however, is not as straightforward as it seems to be, because the correct form of the constitutive relations after *virtual* deformation of *object*s is open to discussion. This point is discussed for each example in Appendix B. The mainstream literature is devoted to *real* deformation of *object*s, e.g. elastomers, exposed to electrostatic fields [41]. Such study does not necessitate distinction between the shares of mechanical and EM forces. This is fortunate because making such a distinction is a heavy burden [42].

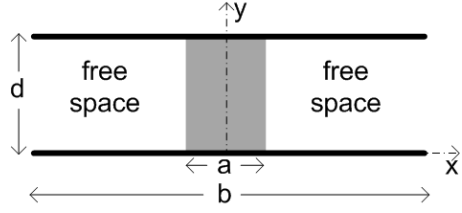

Fig. 2. Two parallel metallic planes partly filled with a linear isotropic *homogeneous object* with a permittivity $\varepsilon$ and a permeability $\mu$. The structure is uniform along the $z$-axis.

### A. Example A

As shown in Fig. 2, the empty space between two parallel metallic planes is partly filled with the *object*. The lower and the upper metallic planes have predetermined charges $Q_0 > 0$ and $-Q_0$ per unit length of the $z$-axis, respectively. Assuming that the electric field is uniform within the space enclosed by the metallic planes, it can be analytically calculated and written as $\vec{E} = \hat{y}\dfrac{Q_0}{a\varepsilon + (b-a)\varepsilon_0}$. The uniformity of the electric field is valid insofar as $b$ is much larger than $d$.

The total EM force exerted on the *object* is zero, but the distribution of the exerted EM force on the *object boundary* hinges upon the choice of the formulation of the EM stress tensor. Both the Ampere-Lorentz force and the Einstein-Laub force exerted on the rightward *boundary* are zero according to Eqs. (12) and (13), whereas use of Eq. (18) (or Eq. (20)) yields $\hat{x}\dfrac{Q_0^2(\varepsilon - \varepsilon_0)}{2(a\varepsilon + (b-a)\varepsilon_0)^2}$ for the exerted Minkowski (or Abraham) force per unit area of the rightward *boundary* of the *object*. Below, their compatibility with special relativity is examined to find the correct result from the *theoretical* point of view.

### 1. Compatibility with special relativity

To provide a relativistic interpretation of Example A, which has been hitherto discussed in the rest frame, we consider a new inertial frame moving at a constant velocity $\hat{z}V$ with respect to the rest frame. In this new frame, the electric field and the magnetic flux density are $\vec{E}' = \hat{y}\dfrac{\gamma Q_0}{a\varepsilon + (b-a)\varepsilon_0}$, and $\vec{B}' = \hat{x}\dfrac{\gamma V Q_0}{c^2(a\varepsilon + (b-a)\varepsilon_0)}$, respectively, where $c$ denotes the speed of light in free space and $\gamma = \dfrac{1}{\sqrt{1 - V^2/c^2}}$. Similarly, the displacement and the magnetic fields outside the *object* (and inside the *object*) are $\vec{D}'_O = \hat{y}\dfrac{\gamma Q_0 \varepsilon_0}{a\varepsilon + (b-a)\varepsilon_0}$ (and $\vec{D}'_I = \hat{y}\dfrac{\gamma Q_0 \varepsilon}{a\varepsilon + (b-a)\varepsilon_0}$) and $\vec{H}'_O = \hat{x}\dfrac{\gamma V Q_0 \varepsilon_0}{a\varepsilon + (b-a)\varepsilon_0}$ (and $\vec{H}'_I = \hat{x}\dfrac{\gamma V Q_0 \varepsilon}{a\varepsilon + (b-a)\varepsilon_0}$), respectively. Therefore, the stress tensors of Ampere-Lorentz, Einstein-Laub, and Minkowski (or Abraham) which appeared in Eqs. (4), (5), and (7) (or (10)), respectively, can now be calculated outside and inside the *object* in the new inertial frame.

It is worth noting that, generally speaking, in a moving frame, each of the displacement field and the magnetic flux density inside a medium other than free space is intricately dependent on both the electric and the magnetic fields, even if the medium is linear and isotropic in the rest frame [43]. However, in this specific example, it turns out that $\vec{D}'_I = \varepsilon \vec{E}'$ and $\vec{B}' = \dfrac{1}{c^2 \varepsilon}\vec{H}'_I$ inside the *object*, and therefore the form of the Minkowski (or Abraham) stress tensor in Eq. (7) (or Eq. (10)) is still applicable.

The forces of Ampere-Lorentz, Einstein-Laub, Minkowski (or Abraham), exerted per unit area of the rightward *boundary* of the *object* in the new inertial frame can now be obtained by using Eqs. (12), (13), and (18) (or (20)), respectively. These per unit area forces are expected to be invariant in the two inertial frames because both the force and the unit area decrease by a factor of $1/\gamma$ in the new inertial frame. Comparison of the results in the new inertial frame and their counterparts in the rest frame, however, demonstrates that the Einstein-Laub force per unit area, in contrast to the others, is *not* invariant in the two inertial frames. This is an inconsistency between the Einstein-Laub formulation and special relativity. The Einstein-Laub force per unit area in the new inertial frame is $\hat{x} \dfrac{Q_0^2(\varepsilon-\varepsilon_0)}{2(a\varepsilon+(b-a)\varepsilon_0)^2} \dfrac{\varepsilon-\varepsilon_0}{\varepsilon_0}(\dfrac{\gamma V}{c})^2$.

*2. Counterpart of Example A in the presence of the free currents*

In Example A, the electric field is parallel to the surface on which the EM force is exerted. The geometry of the counterpart of this example in the presence of the *free* currents is depicted in Fig. 3. It is assumed that in the rest frame, the metallic planes in Fig. 3 have no charge, but carry predetermined surface current densities $\hat{z}J_{s_0}$ and $-\hat{z}J_{s_0}$. It can be easily shown that each of the forces of Ampere-Lorentz and Minkowski (or Abraham) exerted per unit area of the upper *boundary* is invariant in the two inertial frames while the Einstein-Laub force per unit area of the upper *boundary* depends on the velocity of the inertial frame.

**B. Example B**

As another example, the geometrical arrangement in Fig. 3 is considered when the lower and the upper metallic planes have predetermined surface charge densities $\rho_{s_0} > 0$ and $-\rho_{s_0}$, respectively, but carry no current. Therefore, the displacement field within the space enclosed by the metallic planes is written as $\vec{D} = \hat{y}\rho_{s_0}$.

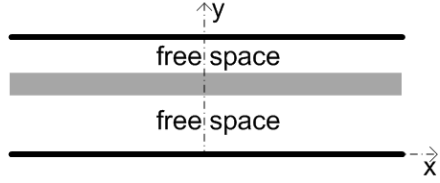

Fig. 3. Two parallel metallic planes with a linear isotropic *homogeneous* slab with a permittivity $\varepsilon$ and a permeability $\mu$. The structure is uniform along both the $z$- and the $x$-axes.

The total EM force exerted on the *object* is zero, but different formulations of the EM stress tensor lead to different distributions of the exerted EM force on the *object boundary*. Use of Eqs. (12), (13), and (18) (or (20)) yield $\hat{y}\dfrac{\rho_{s_0}^2}{2\varepsilon_0}\dfrac{\varepsilon^2-\varepsilon_0^2}{\varepsilon^2}$, $\hat{y}\dfrac{\rho_{s_0}^2}{2\varepsilon_0}(\dfrac{\varepsilon-\varepsilon_0}{\varepsilon})^2$, and $\hat{y}\dfrac{\rho_{s_0}^2}{2\varepsilon_0}\dfrac{\varepsilon-\varepsilon_0}{\varepsilon}$ for the forces of Ampere-Lorentz, Einstein-Laub, and Minkowski (or Abraham), exerted per unit area of the upper *boundary* of the *object*, respectively.

### 1. Compatibility with special relativity

Example B is now reexamined in the new inertial frame moving at the constant velocity $\hat{z}V$ with respect to the rest frame. In the new inertial frame, the displacement and the magnetic fields are $\vec{D}'=\hat{y}\gamma\rho_{s_0}$, and $\vec{H}'=\hat{x}\gamma V\rho_{s_0}$, respectively. Similarly, the electric field and the magnetic flux density outside the *object* (and inside the *object*) are $\vec{E}'_O=\hat{y}\dfrac{\gamma\rho_{s_0}}{\varepsilon_0}$ (and $\vec{E}'_I=\hat{y}\dfrac{\gamma\rho_{s_0}}{\varepsilon}$) and $\vec{B}'_O=\hat{x}\dfrac{\gamma V\rho_{s_0}}{c^2\varepsilon_0}$ (and $\vec{B}'_I=\hat{x}\dfrac{\gamma V\rho_{s_0}}{c^2\varepsilon}$), respectively. The stress tensors in Eqs. (4), (5), and (7) (or (10)) can now be calculated outside and inside the *object* in the new inertial frame. It is worth noting that $\vec{D}'=\varepsilon\vec{E}'_I$ and $\vec{B}'_I=\dfrac{1}{c^2\varepsilon}\vec{H}'$ inside the *object*, and therefore, the form of the Minkowski (or Abraham) stress tensor in Eq. (7) (or Eq. (10)) is still applicable.

The forces of Ampere-Lorentz, Einstein-Laub, and Minkowski (or Abraham) exerted per unit area of the upper *boundary* of the *object* in the new inertial frame are obtained by using Eqs. (12), (13), and (18) (or (20)), respectively. Once again, only the Einstein-Laub force per unit area is *not* invariant in the two inertial frames. It can be shown that the Einstein-Laub force per unit area in the new inertial frame is $\hat{y} \frac{\rho_{s_0}^2}{2\varepsilon_0} (\frac{\varepsilon - \varepsilon_0}{\varepsilon})^2 \gamma^2$. This is incompatible with special relativity because the force exerted on the upper *boundary*, and the unit area of the upper *boundary* both decrease by a factor of $1/\gamma$ in the new inertial frame, and the force per unit area must be invariant in the two inertial frames.

  *2. Counterpart of Example B in the presence of the free currents*

In Example B, the electric field is perpendicular to the surface on which the EM force is exerted. The geometry of the counterpart of this example in the presence of the *free* currents is depicted in Fig. 2. It is assumed that in the rest frame, the metallic planes in Fig. 2 have no charge, but carry predetermined current $I_0$ along the $\hat{z}$ and $-\hat{z}$. Yet again, it can be easily shown that each of the forces of Ampere-Lorentz and Minkowski (or Abraham) exerted per unit area of the rightward *boundary* is invariant in the two inertial frames while the Einstein-Laub force per unit area of the rightward *boundary* depends on the velocity of the inertial frame.

  **V. CONCLUSIONS**

We studied the different EM stress tensors affiliated with the different formulations of the EM force density. Since the mathematical expressions of the stress tensors of Minkowski and Abraham inside a material depend on the constitutive relations of the displacement field and the magnetic flux density, we specifically derived these tensors for three categories of *homogeneous* materials, and show that their forms can be quite different from the well-known forms in the literature.

From a viewpoint based on the law of conservation of momentum in EM interactions, the total EM force exerted on an *object* in a host medium was addressed together with the distribution of the exerted EM force on the *object boundary*. The *object boundary* is in fact the macroscopic representation of the microscopic volume in between the *object* and the host medium. Since the microscopic volume has a non-zero mass, attributing a non-zero force to the *object boundary* is not paradoxical. We argued that finding the exact amount of the exerted EM force on the *object* in a non-free space host medium would require a detailed study of the *object boundary*. Besides, the overall Ampere-Lorentz, Einstein-Laub, Minkowski, and Abraham forces exerted on the *object* in free space when the EM fields are static (or their time averages when the EM fields are time-harmonic) become equal to each other. It should be noted that finding and integrating the contribution of the Helmholtz term in the exerted Minkowski (or Abraham) force is sidestepped by following the presented approach.

By a few simple examples, it was shown that the Einstein-Laub formulation is not necessarily consistent with special relativity. In each of the considered examples, we examined the EM force experienced by a linear isotropic *homogeneous object* inside free space in the presence of the *free* charges (or currents). It should be noted that EM fields in each of the examples are static in both the rest and the moving frames. Therefore, the issue of the hidden momentum is of no consequence in our examples. The same holds true when EM fields are time-harmonic. Otherwise, either hidden entities, e.g. the hidden momentum, come into the play, or the asynchronous formulation [31-33] is to be adopted to circumvent being involved with hidden entities. In the latter approach, the process of synchronously integrating force densities as carried out in section II in the rest frame (the privileged frame in the asynchronous formulation) should be repeated in the moving frame, whereby the stress tensors are obtained. This process is not yet carried out, and thus, can be a good subject for future studies.

**ACKNOWLEDGMENTS**

We thank Dr B. Rejaei for his helpful discussions, and the anonymous reviewer for drawing our attention to the asynchronous formulation of special relativity.

**APPENDIX A**

Here, the Minkowski stress tensors of a few categories of *homogeneous* materials are derived from the Minkowski force density in Eq. (6). According to Maxwell's equations, Eq. (6) can be rewritten as

$$\vec{f}_M = (\nabla \cdot \vec{D})\vec{E} + (\nabla \times \vec{E}) \times \vec{D}$$

$$+ (\nabla \cdot \vec{B})\vec{H} + (\nabla \times \vec{H}) \times \vec{B} - \frac{\partial}{\partial t}(\vec{D} \times \vec{B}). \tag{A.1}$$

To obtain the exerted Minkowski force on a volume $\upsilon$ within a *homogeneous* medium, Eq. (A.1) has to be integrated over the volume. Integration of the fifth term on the right-hand side of Eq. (A.1) yields the volume integral in Eq. (3) with $\vec{G} = \vec{G}_M = \vec{D} \times \vec{B}$, but integration of the other terms is to be carried out. It is sufficient to carry out integration of the first two terms since the next two happen to be their magnetic counterparts. The first two terms can be written as

$$(\nabla \cdot \vec{D})\vec{E} + (\nabla \times \vec{E}) \times \vec{D} = (\nabla \cdot \vec{D})\vec{E}$$

$$+ \hat{x} \sum_{i=x,y,z} D_i (\frac{\partial E_x}{\partial i} - \frac{\partial E_i}{\partial x}) + \hat{y} \sum_{i=x,y,z} D_i (\frac{\partial E_y}{\partial i} - \frac{\partial E_i}{\partial y})$$

$$+ \hat{z} \sum_{i=x,y,z} D_i (\frac{\partial E_z}{\partial i} - \frac{\partial E_i}{\partial z}) = \nabla \cdot (\vec{D}\vec{E}) - \hat{x} \sum_{i=x,y,z} D_i \frac{\partial E_i}{\partial x}$$

$$-\hat{y} \sum_{i=x,y,z} D_i \frac{\partial E_i}{\partial y} - \hat{z} \sum_{i=x,y,z} D_i \frac{\partial E_i}{\partial z}. \tag{A.2}$$

The last three terms on the right-hand side of Eq. (A.2) are behind the fact that the form of the Minkowski stress tensor is influenced by the constitutive relation of $\vec{D}$. Thanks to the similarity among these three terms, we consider the $\hat{x}$ component only. As the first category, we assume that the constitutive relation of $\vec{D}$ is linear and reciprocal, i.e. $\vec{D} = \vec{\vec{\varepsilon}}\vec{E}$ with a symmetric tensor $\vec{\vec{\varepsilon}}$:

$$\hat{x} \sum_{i=x,y,z} D_i \frac{\partial E_i}{\partial x} = \frac{\hat{x}}{2} \frac{\partial (\vec{D} \cdot \vec{E})}{\partial x}. \tag{A.3}$$

From Eqs. (A.2), (A.3), and its counterparts along the $y$- and the $z$-axes, it becomes evident that

$$(\nabla \cdot \vec{D})\vec{E} + (\nabla \times \vec{E}) \times \vec{D} = \nabla \cdot (\vec{D}\vec{E}) - \frac{1}{2} \nabla \cdot (\vec{I}(\vec{D} \cdot \vec{E})), \tag{A.4}$$

where $\vec{I}$ is the identity tensor. Therefore, if the constitutive relation of $\vec{D}$ is linear and reciprocal, the contribution of the electric and the displacement fields to the form of the Minkowski stress tensor is $\vec{D}\vec{E} - \frac{1}{2}\vec{I}(\vec{D} \cdot \vec{E})$. It can be easily shown that the same is true if the material is linear, isotropic, and dispersive, i.e. $\vec{D}(t) = \varepsilon(t) * \vec{E}(t)$ when $*$ denotes convolution.

As another category, we assume that the constitutive relation of $\vec{D}$ is nonlinear with a permanent electric polarization $\vec{P}_0$, i.e. $\vec{D} = \varepsilon_0 \vec{E} + \vec{P}_0$:

$$\hat{x} \sum_{i=x,y,z} D_i \frac{\partial E_i}{\partial x} = \frac{\hat{x}}{2} \frac{\partial (\varepsilon_0 \vec{E} \cdot \vec{E} + 2\vec{P}_0 \cdot \vec{E})}{\partial x}. \tag{A.5}$$

Now from Eqs. (A.2), (A.5), and its counterparts along the $y$- and the $z$-axes, it becomes evident that the contribution of the electric and the displacement fields to the form of the Minkowski stress tensor when the constitutive relation of $\vec{D}$ is nonlinear with a permanent electric polarization $\vec{P}_0$ is $\vec{D}\vec{E} - \frac{1}{2}\vec{I}(\varepsilon_0 \vec{E} \cdot \vec{E} + 2\vec{P}_0 \cdot \vec{E})$.

**APPENDIX B**

To obtain the exerted EM force on the rightward *boundary* of the *object* in Example A by the method of virtual work, the change in the electric energy has to be calculated when the rightward *boundary* is infinitesimally displaced along the $x$-axis, while the rest of the *object boundary* remains still. To this end, it is essential to first specify how the constitutive relation of $\vec{D}$ inside the *object* varies with infinitesimal expansion/contraction of the *object* along the $x$-axis. This dependence of the result of the method of virtual work on the constitutive relations inside the *virtually expanded/contracted object* renders the resultant EM force debatable, because the EM force is expected to be influenced only by the constitutive relations when the *object* is in its original state.

The exerted EM force per unit area of the rightward *boundary* obtained by the method of virtual work is in agreement with the result of the formulations of Ampere-Lorentz and Einstein-Laub if $\varepsilon = \varepsilon_0(1 + a/L)$, where $L$ is the *object*'s thickness which *virtually* expands/contracts, namely the *object*'s thickness along the $x$-axis in this example, and $a$ is a constant coefficient whose physical dimension is *length*. The result of the method of virtual work is in agreement with the result of the Minkowski (or Abraham) formulation if $\varepsilon$ is assumed to remain unchanged when the *object virtually* expands/contracts.

To apply the method of virtual work to Example B, the change in the electric energy has to be calculated when the upper *boundary* is infinitesimally displaced along the $y$-axis, while the rest of the *object boundary* remains unchanged. The exerted EM force per unit area of the upper *boundary* obtained by the method of virtual work is in agreement with the result of the Ampere-Lorentz formulation if $\varepsilon = \varepsilon_0(1+bL)$, where $L$ is the *object*'s thickness which *virtually* expands/contracts, namely the *object*'s thickness along the $y$-axis in this example, and $b$ is a constant coefficient whose physical dimension is *1/length*. The result of the method of virtual work is in agreement with the Einstein-Laub formulation if $\varepsilon = \varepsilon_0(1+a/L)$, and with the result of the Minkowski (or Abraham) formulation if $\varepsilon$ does not vary with *virtual* expansion/contraction of the *object*.

[*] jazayeri@ee.sharif.edu

[‡] mehrany@sharif.edu